\documentclass[american,superscriptaddress,aps,reprint,showpacs]{revtex4-1}
\usepackage[latin9]{inputenc}
\usepackage{textcomp}
\usepackage{amsmath}
\usepackage{amssymb}
\usepackage{graphicx}
\usepackage{babel}
\begin{document}

\title{Measurement of Anticipative Power of a Retina by Predictive Information}

\author{Kevin Sean Chen}

\affiliation{Institute of Physics, Academia Sinica, Taipei, Taiwan 115, Republic
of China}

\affiliation{Department of Life Science, National Taiwan University, Taipei, Taiwan,
Republic of China}

\author{Chun-Chung Chen}
\email{cjj@phys.sinica.edu.tw}

\selectlanguage{american}%

\affiliation{Institute of Physics, Academia Sinica, Taipei, Taiwan 115, Republic
of China}

\author{C. K. Chan}
\email{ckchan@gate.sinica.edu.tw}

\selectlanguage{american}%

\affiliation{Institute of Physics, Academia Sinica, Taipei, Taiwan 115, Republic
of China}

\affiliation{Department of Physics and Center for Complex Systems, National Central
University, Chungli, Taiwan 320, Republic of China}

\date{November 10, 2016}
\begin{abstract}
The predictive properties of a retina are studied by measuring the
mutual information (MI) between its stimulation and the corresponding
firing rates while it is being probed by a train of short pulses with
stochastic intervals. Features of the measured MI at various time
shifts between the stimulation and the response are used to characterize
the predictive properties of the retina. By varying the statistical
properties of the pulse train, our experiments show that a retina
has the ability to predict future events of the stimulation if the
information rate of the stimulation is low enough. Also, this predictive
property of the retina occurs at a time scale similar to the well
established anticipative phenomenon of omitted stimulus response in
a retina. Furthermore, a retina can make use of its predictive ability
to distinguish between time series created by an Ornstein\textendash Uhlenbeck
and a hidden Markovian process.
\end{abstract}

\maketitle

\section{Introduction}

The ability to predict or anticipate future events is crucial for
the survival of animals. Predicting dynamical inputs can compensate
the latency during information transfer and provide predictive information
for learning and behavior \cite{berry_retina_2011,berry_anticipation_1999,hosoya_dynamic_2005,leonardo_nonlinear_2013,bialek_predictability_2001}.
In 2007, Schwartz \cite{schwartz_detection_2007,schwartz_sophisticated_2008}
et al reported that there will be spontaneous responses from the ganglion
cells in the retina of salamander and mice after a periodic light
stimulation is abruptly stopped; with the latency of this spontaneous
response being linearly related to the period of the stopped stimulation.
In other words, the retina seems to anticipate when the next upcoming
pulse should have occurred and produce a response if the upcoming
pulse is missing. This timed response for the omitted pulse from the
retina is known as omitted stimulus response (OSR). Phenomena similar
to the OSR have also been reported for induced ocular motor behavior
under periodic light stimuli in zebra fish larvae \cite{sumbre_entrained_2008}
and growth of slime mold under periodic alternation of moisture or
temperature \cite{saigusa_amoebae_2008}.

Ideally, one would like to quantify and model the predictive properties
of a retina. Although the phenomenon of OSR has been discovered for
more than 10 years, it is still not clear how to relate OSR to the
predictive properties of the retina. In OSR, information of the stimulation
is apparently coded into the timing of the pulses. However, when there
are fluctuations in the inter-pulse intervals of the stimulation,
it is difficult to identify or even produce OSR. Therefore, it is
not feasible to make use of OSR in inferring the predictive properties
of a retina for the general cases of a non-periodic stimulation which
should contain much more information than a purely periodic one. Bialek
and Tishby have introduced the idea of predictive information (PI)
based on the statistical properties of the input and output signal
of a data processing system \cite{rubin_representation_2016,bialek_predictive_1999}.
Recently, this idea has been applied successfully to describe the
response of a retina to a stimulation in the form of a stochastic
moving bar by computing the mutual information, $I_{m}\left(\delta t\right)$,
between the input and output as a function of time shift $\delta t$
between the two signals \cite{palmer_predictive_2015}. Intuitively,
the form of $I_{m}\left(\delta t\right)$ should be determined by
the predictive dynamics of the retina. However, it is still not clear
what kind of information one can extract from $I_{m}\left(\delta t\right)$.

In this work, we report our experimental results in quantifying the
predictive properties of a retina by using the PI method mentioned
above. With a retina plated on top of a multi-electrode array (MEA)
probed by stochastic light pulses, $I_{m}\left(\delta t\right)$ is
measured as a function of the properties of the light pulses; namely
its mean inter-pulse interval ($\left\langle \tau\right\rangle $)
and correlation time ($\tau_{cor}$). Our main finding is that the
location of the peak of $I_{m}\left(\delta t\right)$ can be shifted
from $-\delta t$ to $+\delta t$ by an increase of $\tau_{cor}$;
suggesting that retina has the ability to predict (with some uncertainties)
future events in the stimulation when the stimulation is regular enough.
However, this ability of prediction can only be observed when $\left\langle \tau\right\rangle $
is in the range of 100 ms \textless{} $\left\langle \tau\right\rangle $
\textless{} 200 ms; similar to that of the OSR phenomenon mentioned
above measured in bullfrog retinas. Furthermore, this predictive property
of a retina can be used to distinguish the signals generated from
an Ornstein\textendash Uhlenbeck (OU) and an hidden Markovian (HMM)
process; with the signal from the HMM process being identified as
more predictable by the retina.

\section{Materials and Methods}

Retinas used in the experiments are obtained under dim red light from
bullfrogs which were dark adapted for 1 hour before dissection. Our
sample is consisted of a piece of retina fixed on the 60-channel multi-electrode
array (MEA, 200 \textmu m inter-electrode distance with 10 \textmu m
electrode diameter) by a permeable membrane and perfused with oxygenated
Ringer's solution. Each retina preparation can last for 6-8 hours.
Stimulations to the retina is in the form of a train of stochastic
light pulses (pulse duration = 50 ms) generated from a LED (peak of
wavelength = 560 nm, intensity = 5 cd/m$^{2}$) which illuminates
the whole retina through a projection lens. The interval between pulses
is controlled by a computer to produce a train of pulses with different
characteristics which will be described in details below. Responses
from the retinas are recorded at 20 kHz through the local field potentials
at the 60 electrodes of the MEA. Spike sorting is performed through
the Offline Sorter software. Signals with ambiguous or multiple waveforms
are discarded. Firing rates are calculated as the number of spikes
identified within a 5 ms bin. Our experiments consist of recording
responses of the retina for stimulations with different characteristics.
The protocol is to present each set of stimuli continuously for 5
min in a random order, and the inter-experiment resting time is 2\textendash 3
min. For OSR measurements, the same stimuli constituted of 20 pulses
are repeated for 10\textendash 20 trials with 3\textendash 5 sec inter-trial
resting time. All the experiments are carried out in a dark room with
temperature around 25 \textdegree C. In the results reported below,
over ten retina samples are used and at least three retina samples
(on average 10\textendash 20 waveforms sorted from each sample) were
used to verify each experimental results.

\section{Results}

\begin{figure}
\begin{centering}
\includegraphics[width=0.9\columnwidth]{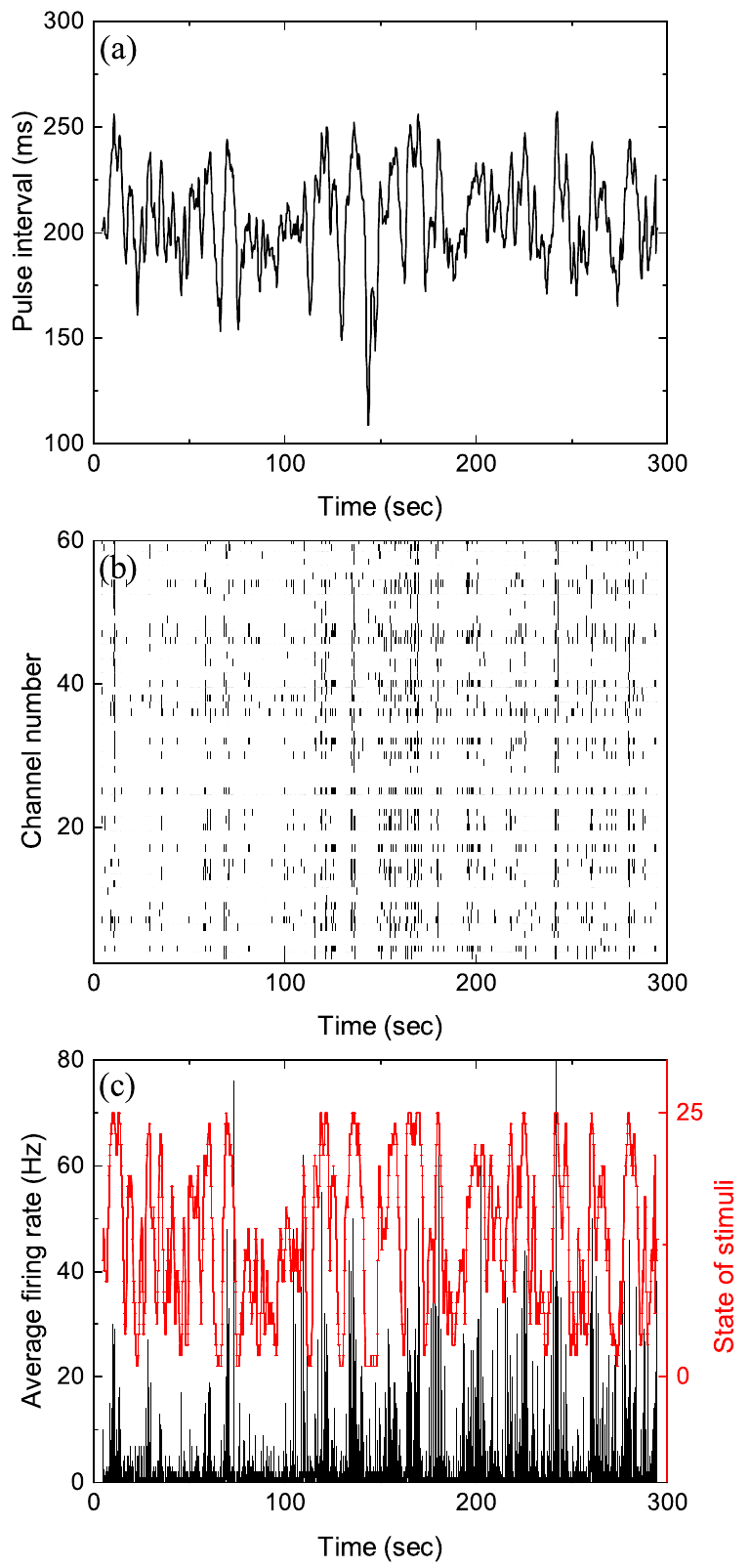} 
\par\end{centering}
\caption{Stochastic pulse intervals and the induced retinal firing patterns.
(a) Time series of pulse intervals generated by the iteration formula;
$\langle\tau\rangle=200$ ms, $\sqrt{\langle(\tau-\langle\tau\rangle)^{2}\rangle}=20$
ms and $\tau_{cor}=4$ s. (b) Raster plot showing firing timestamps
from 60 channels under the input shown in (a). (c) Average firing
rate of the population recorded in (b). To calculate mutual information,
the stimuli shown in (a) with varying pulse intervals are defined
as equally distributed 25 states shown in red.}
\label{fig1:stochastic pulses} 
\end{figure}

Figure~\ref{fig1:stochastic pulses}a shows inter-pulse-interval
($\tau$) of the stochastic pulse train used in the experiment as
a function of time (with a discrete time step of 5 ms). The pulse
train is characterized by three parameters; namely the mean inter-pulse
interval ($\langle\tau\rangle$), the correlation time between inter-pulse
intervals ($\tau_{cor}$) and the standard deviations of $\tau$.
This stimulation series is generated by following the idea of Palmer
et al \cite{palmer_predictive_2015} which is associated with a damped
harmonic oscillator driven by noise, with the $i^{th}$ intervals
being generated as:
\begin{eqnarray}
\tau_{i+1} & = & \tau_{i}+v_{i}\Delta\label{eq:schemeP}\\
v_{i+1} & = & \left(1-\Gamma\right)v_{i}-\omega^{2}\tau_{i}\Delta+\xi_{i}\sqrt{D\Delta}\label{eq:hidden-v}
\end{eqnarray}
where \emph{v} is the rate of change of $\tau$, $\xi$ a Gaussian
noise with zero mean with amplitude $D=1$. The iteration step size
$\Delta$ is fixed at 1/60 in the iteration. Note that $\Gamma$/2$\omega$
is kept at 1.06 so that the system is slightly over-damped. To generate
the stimulations, the series $\left\{ \tau_{i}\right\} $ ($\equiv\left\{ \tau_{1},\tau_{2},\ldots\right\} $)
is first created by the iteration of Eq.~\eqref{eq:schemeP} and
Eq.~\eqref{eq:hidden-v}. Then, the standard deviations of $\left\{ \tau_{i}\right\} $
is rescaled to have a fixed value of 20 ms and an offset is added
to $\left\{ \tau_{i}\right\} $ to obtain the desired mean $\langle\tau\rangle$.
With this method, the correlation of $\left\{ \tau_{i}\right\} $
is not only controlled by $\Gamma$, the rescaling of its standard
deviation and the addition of offset all affect the correlation time
of the series. The correlation time of the resultant stimulation must
then be measured by computing its autocorrelation function. Note that
when $\tau_{cor}$ tends to $\infty$, we will recover the periodic
stimulation in OSR. With this stochastic pulse train, we can stimulate
the retina by temporal patterns with continuous adjustable $\langle\tau\rangle$
and $\tau_{cor}$. During each experiment reported below, such a pulse
train is presented to the retina for 5 min. Figure~\ref{fig1:stochastic pulses}b
is the raster plot of the firings of the retina recorded by the MEA
while Fig.~\ref{fig1:stochastic pulses}c shows the average firing
rate obtained from Fig.~\ref{fig1:stochastic pulses}b. 

\begin{figure}
\begin{centering}
\includegraphics[width=0.9\columnwidth]{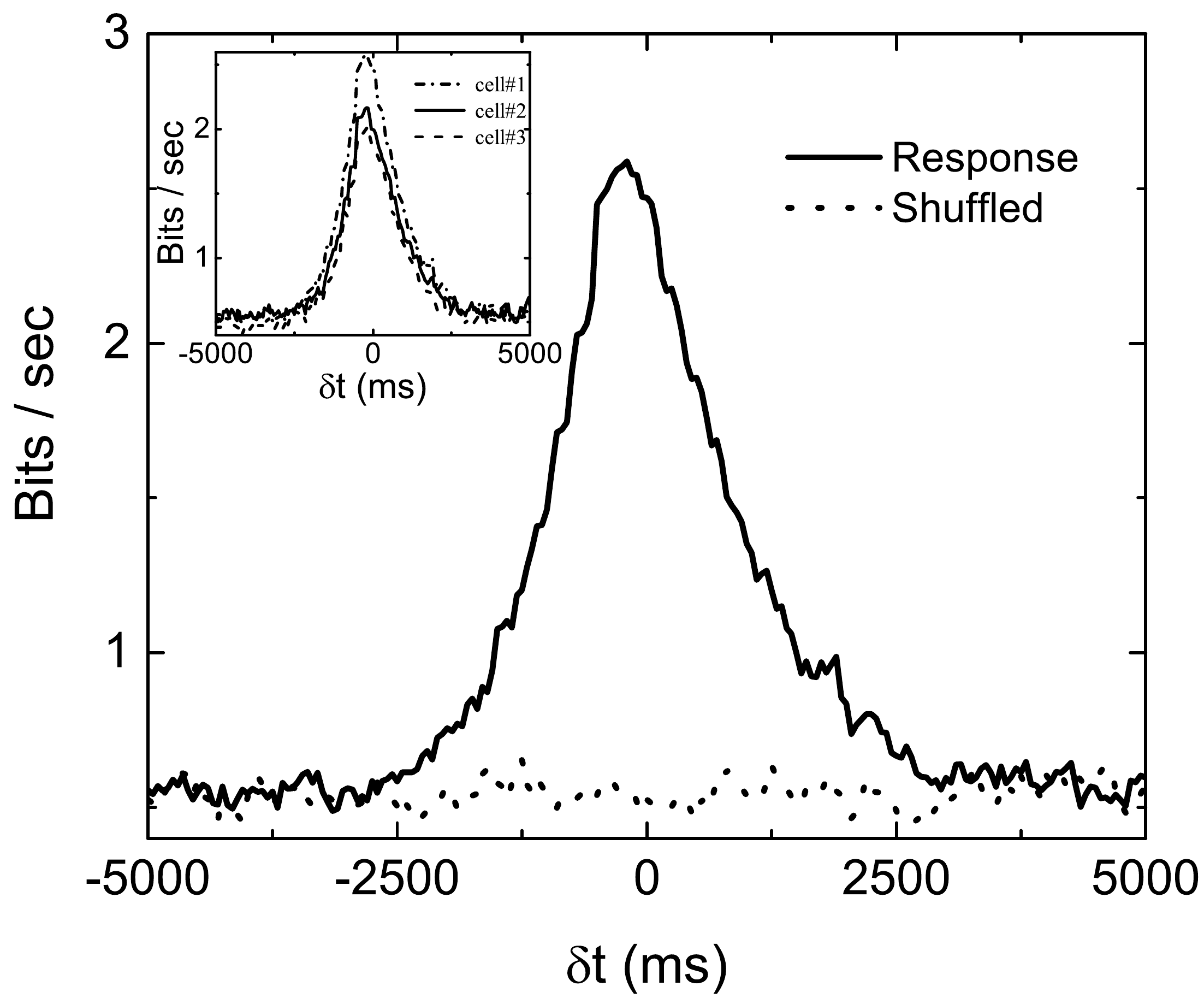} 
\par\end{centering}
\caption{An example of measured $I_{m}\left(\delta t\right)$ with stimulation
shown in Fig.~\ref{fig1:stochastic pulses}a. $I_{m}\left(\delta t\right)$
computed from shuffled data is also shown to serve as a base line.
Three different $I_{m}\left(\delta t\right)$ obtained from three
sorted signals in the same experiment are shown in the inset to demonstrate
the variability of the data.}
\label{fig2:predictive information curve} 
\end{figure}

Mutual information at different time lag ($\delta t$) between the
stimulation (Fig.~\ref{fig1:stochastic pulses}a) and response (Fig.~\ref{fig1:stochastic pulses}b)
can then be calculated by using appropriate binning for the stimulation
and response into discrete states. Figure~\ref{fig2:predictive information curve}
is the computed MI between stimulation and response from sorted firing
waveforms in Fig.~\ref{fig1:stochastic pulses}. The stimulation
was binned into 25 equally distributed states ($S=\left\{ s_{1},s_{2},\ldots,s_{25}\right\} $)
while the number of spikes in one time window is used as the state
index for the response ($R=\left\{ r_{1},r_{2},\ldots\right\} $).
The number of states for the response is then the maximum number of
spikes for each channel within the time window. The maximal states
within 50 ms is on average 10\textendash 15 spikes. The mutual information
at time shift $\delta t$ is then given by: 
\begin{equation}
I_{m}\left(S,R,\delta t\right)=\sum_{i}\sum_{j}p\left(s_{i},r_{j-k}\right)\log_{2}\frac{p\left(s_{i},r_{j-k}\right)}{p\left(s_{i}\right)p\left(r_{j-k}\right)}
\end{equation}
where $p\left(x_{i}\right)$ is the probability of having a state
$x_{i}$ and $p\left(s_{i},r_{j}\right)$ is the joint probability
of the state $\left(s_{i},r_{j}\right)$. Note that the difference
in state index $j-k$ denotes a shift in time of $\delta t$. It can
be seen from Fig.~\ref{fig2:predictive information curve}a that
the $I_{m}\left(S,R,\delta t\right)$ has a peak located at negative
$\delta t$ and it is non-zero for $\delta t>0$. The location of
the peak at negative $\delta t$ indicate that maximum information
shared between $S$ and $R$ when $R$ is lagged behind $S$; confirming
our intuition that the retina takes some time to reflect/process the
information contained in $S$ in producing $R$. 

\noindent Similar to the finding of \cite{palmer_predictive_2015},
the nonzero value of $I_{m}\left(S,R,\delta t\right)$ in Fig.~\ref{fig2:predictive information curve}a
for $\delta t>0$ indicates that the firing patterns in retina provides
predictive information for the future events in $S\left(t\right)$
from its history. In fact, $I_{m}\left(S,R,\delta t>0\right)$ is
termed predictive information in \cite{bialek_predictive_1999}. Also,
it can be seen that $I_{m}\left(S,R,\delta t\right)$ can be nonzero
extending into quite large $\delta t$; much longer than the correlation
time of $S$. This last non-physical property of the measured $I_{m}$
originates from the fact we are computing $I_{m}$ from a finite time
series. In order to find the baseline of our experimental $I_{m}$,
either states of stimuli or firing patterns were randomly shuffled.
The MI curve for shuffled data fluctuates at a nonzero and aligns
with measurements at large $\delta t$, showing $I_{m_{0}}$ as a
baseline due to finite data in Fig.~\ref{fig2:predictive information curve}b.
Note that $I_{m_{0}}$ must be obtained for each firing patterns under
different stimulations. $I_{m}$ reported below are all corrected
as: $I_{m}^{*}=I_{m}-I_{m_{0}}$. 

\begin{figure}
\begin{centering}
\includegraphics[width=0.9\columnwidth]{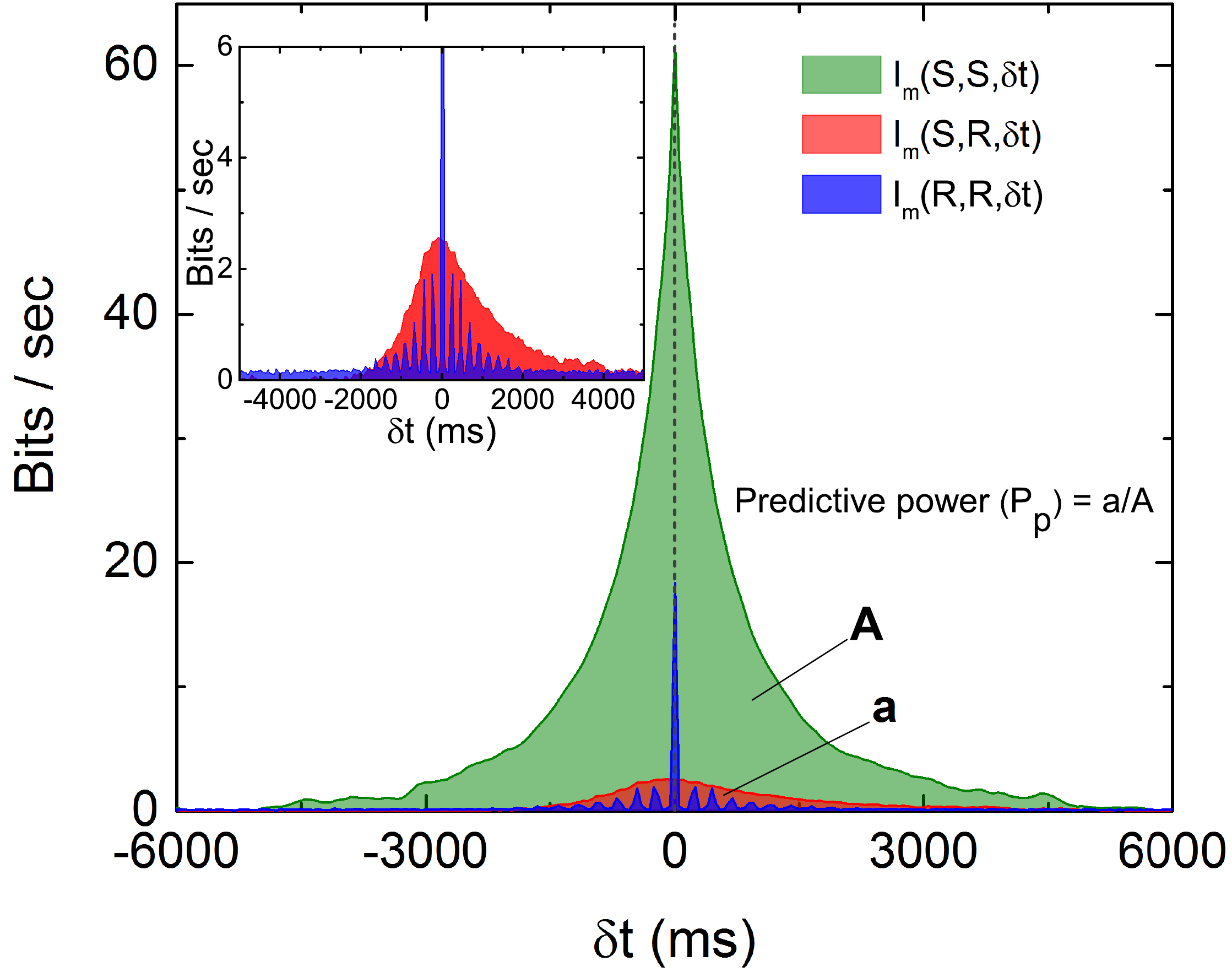} 
\par\end{centering}
\caption{Comparison of the three $I_{m}\left(\delta t\right)$ as described
in the text and the definition of predictive power ($P_{p}$). Note
that both $I\left(S,S,\delta t\right)$ and $I\left(R,R,\delta t\right)$
are symmetric about their respective peaks but $I\left(S,R,\delta t\right)$
is not symmetric (inset). The oscillation observed in $I\left(R,R,\delta t\right)$
is caused by the quasi-periodicity of the stimulation light pulses. }
\label{fig3:comparison} 
\end{figure}

To visualize how much information is being shared between $S$ and
$R$, Fig.~\ref{fig3:comparison} is a comparison of $I_{m}\left(S,S,\delta t\right)$,
$I_{m}\left(R,R,\delta t\right)$ and $I_{m}\left(S,R,\delta t\right)$
from data displayed in Figs.~\ref{fig1:stochastic pulses} and \ref{fig2:predictive information curve}.
It can be seen that only a very small percentage of the information
is being shared by $S$ and $R$. To quantify the amount of predictive
information extracted by the retina, we have defined the predicting
power based on measured $I_{m}$ as the ratio between the two areas
in Fig.~\ref{fig3:comparison} as $P_{p}\left(S,R\right)=a/A$, where
$A$ and $a$ are the area under the curves $I_{m}\left(S,S,\delta t\right)$
and $I_{m}\left(S,R,\delta t\right)$ at positive $\delta t$ respectively.
This definition satisfies the intuitive notion that $P_{p}\left(S,S\right)$
or $P_{p}\left(R,R\right)$ equals to 1 and will allow the comparison
of predictive information between different experiments. A remarkable
feature of Fig.~\ref{fig3:comparison} is that while both $I_{m}\left(S,S,\delta t\right)$
and $I_{m}\left(R,R,\delta t\right)$ decay symmetrically about $\delta t=0$,
$I_{m}\left(S,R,\delta t\right)$ seems to decay slower for $\delta t>0$.
Since both $R$ and $S$ are symmetric with respect to time lag, the
asymmetry of $I_{m}\left(S,R,\delta t\right)$ possibly comes from
the anticipative nature of the retina dynamics in generating $R$.

\begin{figure}
\begin{centering}
\includegraphics[width=0.9\columnwidth]{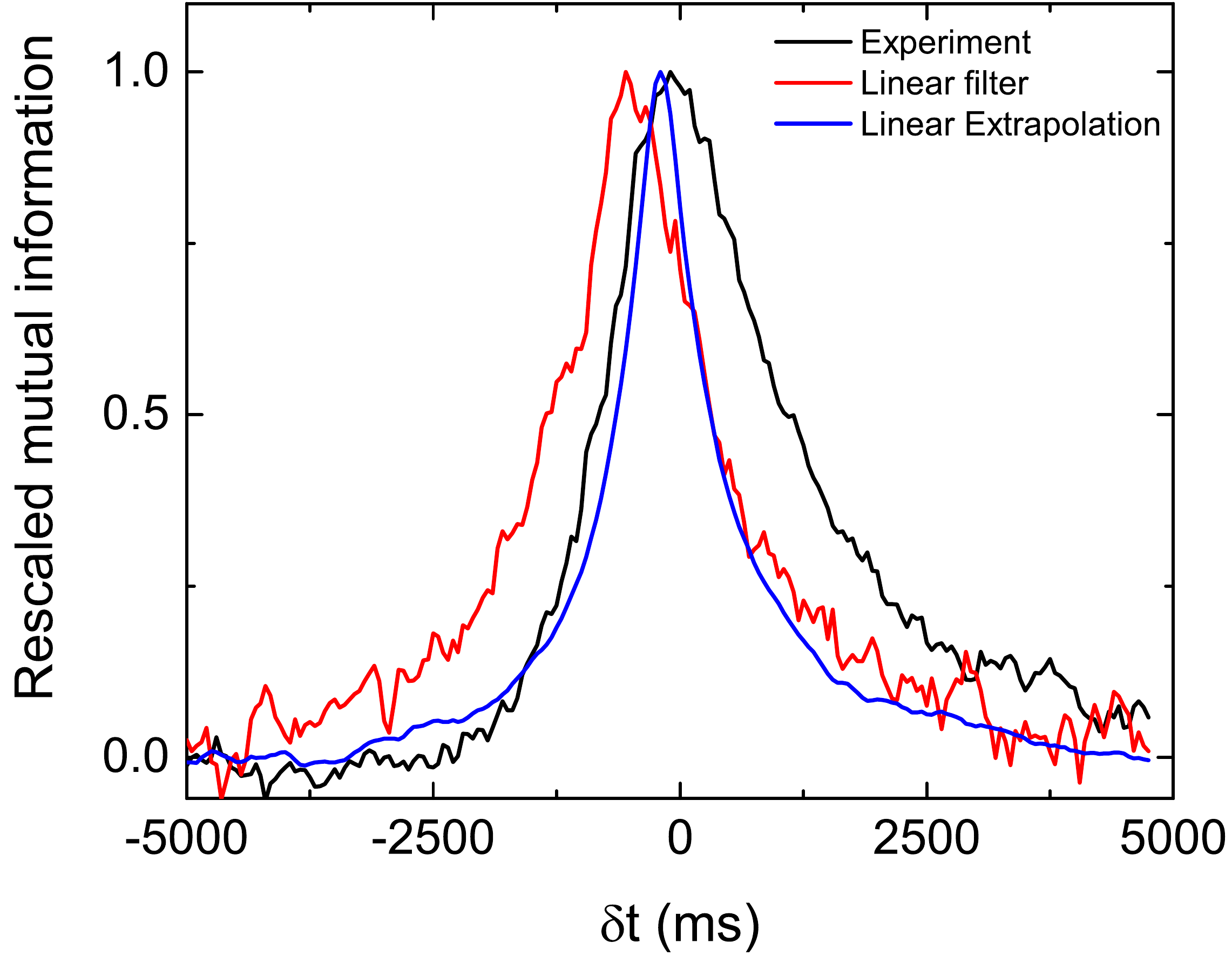} 
\par\end{centering}
\caption{Comparison of $I_{m}\left(\delta t\right)$ with two simulated response.
The experimental result (with parameters the same as those in Fig.~\ref{fig1:stochastic pulses})
is shown in black. The red curve is obtained from the simulation convoluted
with spike trigger average of the retina. The blue curve is simulated
by linear extrapolation with a time window of 10 steps. These $I_{m}\left(\delta t\right)$s
are normalized by their peak values for the ease of comparison. See
text for details.}
\label{fig4:comparison2} 
\end{figure}

To test the idea that asymmetry of $I_{m}\left(S,R,\delta t\right)$
is a manifestation of the predictive nature of the retina, two sets
of $R$ are created artificially. First, the standard method to capture
response of retina was applied, convolving the temporal spike trigger
average (STA) obtained under random flicking stimuli with the stochastic
pulse provided in experiment. The result fails to capture the asymmetry
observed in experiment and over estimates the response delay. Alternatively,
we have simulated a simple anticipative response from the stimulations
$\left\{ s_{i}\right\} $ as $\left\{ r_{i}\right\} $ with $r_{i+1}=s_{i}+v_{i}\delta t$
where $v_{i}$ is the estimated ``velocity'' of the signal based
on its $N$-step history $\left\{ s_{i-N},s_{i-N+1},\ldots,s_{i}\right\} $.
In other words, we are using linear extrapolation of $s_{i}$ to produce
$r_{i+1}$. Figure~\ref{fig4:comparison2} is the computed $I_{m}\left(\delta t\right)$
between the stimulations and their linear extrapolated responses together
with the experimentally measured $I_{m}\left(\delta t\right)$. It
can be seen that this simulated $I_{m}\left(\delta t\right)$ captures
two essential features of the experimental measurements. First, the
peak of the PI curve is not located at zero but at negative time lag.
Second, the decay in $I_{m}\left(\delta t\right)$ is slower for $\delta t>0$.
Obviously, the location of the peak of the $I_{m}\left(\delta t\right)$
curve in Fig.~\ref{fig4:comparison2} is determined by the number
of extrapolation. If we are extrapolating $n$-steps with $r_{i+n}=f\left(s_{i},s_{i-1},\ldots\right)$
for some function $f$, then the peak should be located at minus $n$
time lags. 

\begin{figure}
\begin{centering}
\includegraphics[width=0.9\columnwidth]{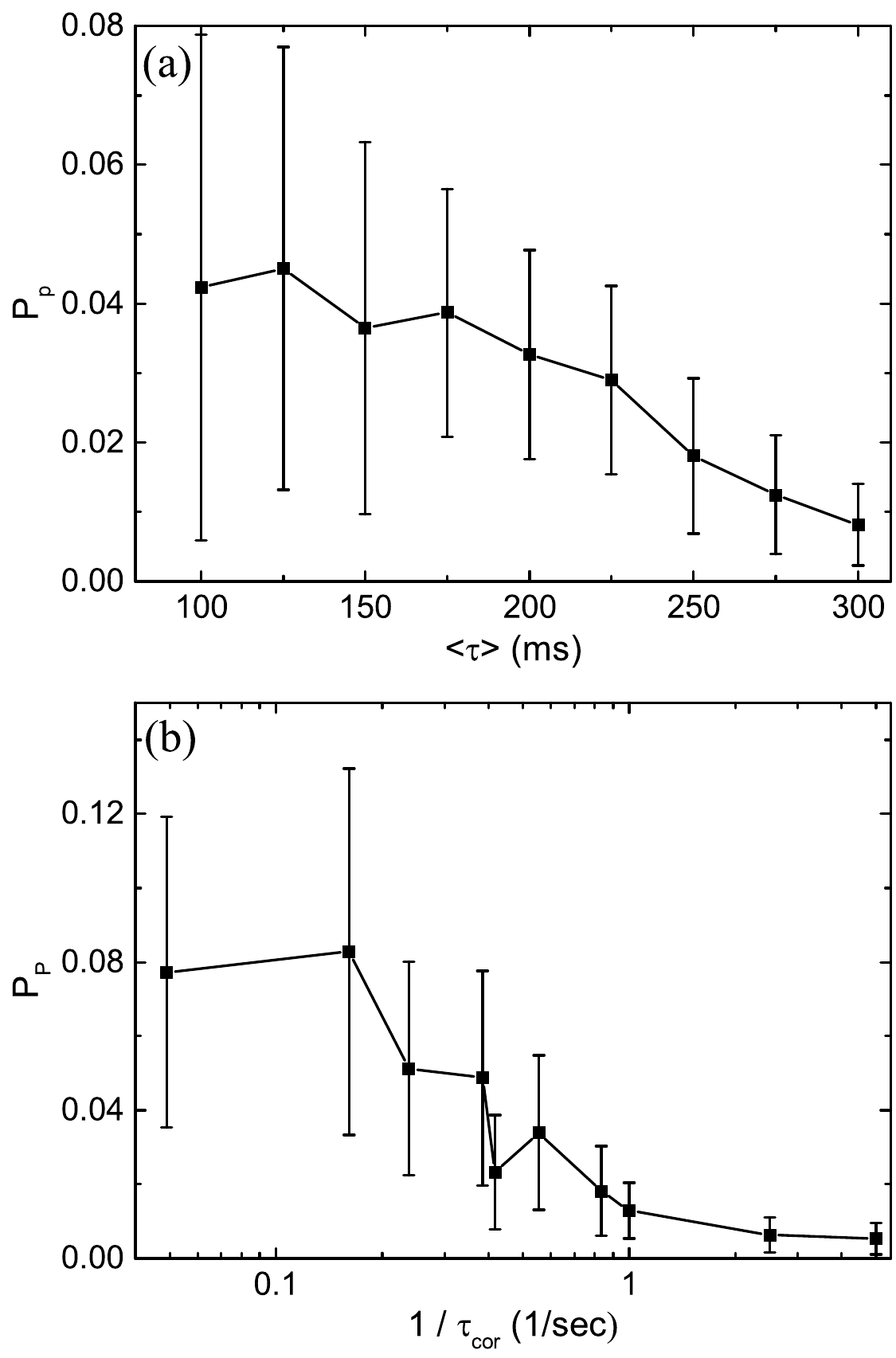} 
\par\end{centering}
\caption{Predictive power ($P_{p}$) depends on the statistical properties
of the simulation light pulses. (a) Measured $P_{p}$ at various $\langle\tau\rangle$
with $\tau$ being fixed to correlate in 20 steps for each value of
$\langle\tau\rangle$. (b) Measured $P_{p}$ as a function of $1/\tau_{cor}$
with $\langle\tau\rangle$ fixed at 200 ms. The results are obtained
from the same retina, and error bar indicates the deviation between
19 sorted signals.}
\label{fig5:predictive power} 
\end{figure}

With the normalization introduced in Fig.~\ref{fig3:comparison},
we can compare the predictive power ($P_{p}$) for stimulations with
various $\langle\tau\rangle$ and $\tau_{cor}$. Figure~\ref{fig5:predictive power}
shows the measured dependence of $P_{p}$ on $\langle\tau\rangle$
and $\tau_{cor}$ by experiments similar to those shown in Fig.~\ref{fig3:comparison}.
Results shown in Fig.~\ref{fig5:predictive power} are obtained from
one single retina. The $P_{p}$ is measured for each channels of the
MEA and error bars are obtained from the spread of these measured
values. With fixed $\tau_{cor}=4$ s, it can be seen from Fig.~\ref{fig5:predictive power}a,
$P_{p}$ falls off to a very small value around $\langle\tau\rangle=200$\textendash $250$
ms. Note that a time scale of 200 ms is also the upper limit for a
periodic stimulation to produce OSR in the bullfrog retina. Figure~\ref{fig5:predictive power}b
shows $P_{p}$ under stimuli with different $\tau_{cor}$ when $\langle\tau\rangle$
fixed at $200$ ms. Note that the data is plotted in the inverse of
$\tau_{cor}$. The idea is that the amount of information encoded
into time series should increase with the inverse of its correlation
time because an purely periodic signal (infinite correlation time)
will not contain any information. With this interpretation, Fig.~\ref{fig5:predictive power}b
indicates that the predictive power of the retina seems to be at its
maximum when the information content of the stimulation is low and
tends to its minimum when the information content is high. The characteristic
time scale (halfway between the max and the min) determined from Fig.~\ref{fig5:predictive power}
is when $\tau_{cor}\approx2.5$ s. 

\begin{figure}
\begin{centering}
\includegraphics[width=0.9\columnwidth]{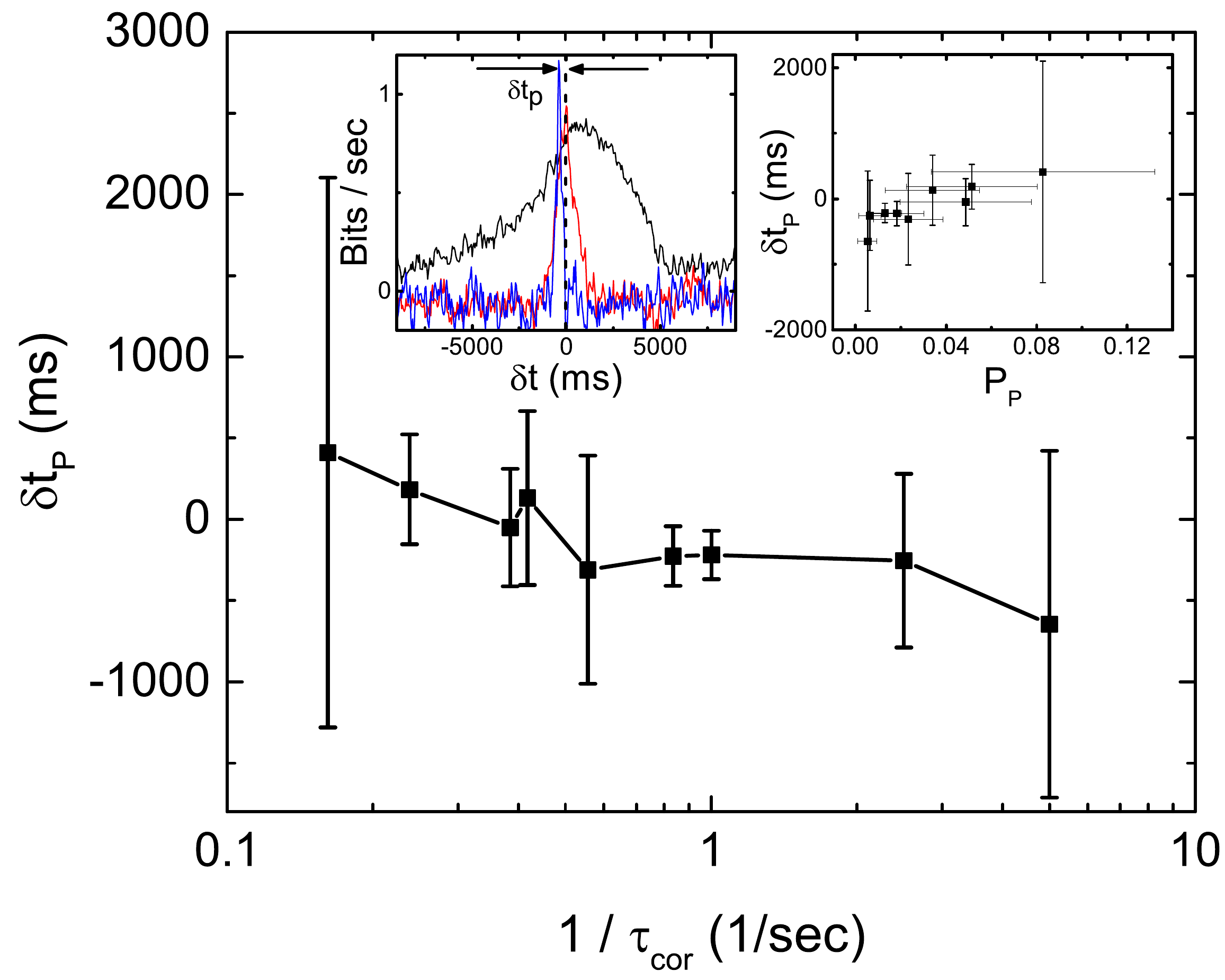} 
\par\end{centering}
\caption{Latency to peak ($\delta t_{p}$) of $I_{m}\left(\delta t\right)$
as a function of $\tau_{cor}$ obtained from 19 sorted signals in
the same retina. The left inset shows the definition of $\delta t_{p}$
and the measured $I_{m}\left(\delta t\right)$ with $\tau_{cor}$
= 0.2 (blue), 4 (red), 7 s (black). Right inset shows the relation
between $\delta t_{p}$ and $P_{p}$ calculated from the same data}
\label{fig6:latency} 
\end{figure}

One interesting feature of the measured $I_{m}$ during our scan of
$\tau_{cor}$ at fixed $\langle\tau\rangle$ is that the peak location
of the $I_{m}$ shifted from negative $\delta t$ to positive $\delta t$
as $\tau_{cor}$ is increased. Figure~\ref{fig6:latency} shows the
dependence of $\delta t_{p}$ as a function of $\tau_{cor}^{-1}$
where $\delta t_{p}$ is the distance of the peak location of $I_{m}$
from the line of $\delta t=0$. The inset of Fig.~\ref{fig6:latency}
shows the definition of peak location ($\delta t_{p}$) and the forms
of $I_{m}\left(\delta t\right)$ for $\tau_{cor}=0.2$, $4.0$, and
$7.0$ s. At first sight, one might expect $\delta t_{p}$ to be always
negative because it will always take time for stimulations just to
propagate through the different layers and synapses of the retina.
That will be true if the retina is just a passive filter. The fact
that $\delta t_{p}$ can be shifted to positive indicates that the
retina is actively predicting the future events of the stimulations.
This interpretation is consistent with the result of Fig.~\ref{fig5:predictive power}b
which suggests that prediction is possible only when the input signal
is regular enough. 

\section{Discussions}

Although the periodic input used in OSR and the stochastic pulses
used in this study seem to be quite different, the periodic pulses
are in fact a limiting case of the stochastic pulses when the correlation
time of the inter-pulse interval becomes infinite. With this consideration,
one can think of the periodic pulses used in the phenomenon of OSR
as the carrier of information very much like the carrier wave in an
FM radio signal and the information is being encoded into the deviations
(fluctuations) from the carrier period. Therefore, the stochastic
pulses (with a fixed mean period) used in our experiments are then
encoding information in its deviations from the mean. The amount of
information encoded can then be characterized by the correlation time;
the longer the $\tau_{cor}$, the less the amount of encoded information.
With a periodic stimulation (infinite correlation time), there is
no information encoded. In fact, this carrier wave picture is supported
by our finding that both OSR and $\langle\tau\rangle$ for optimal
prediction have the same time scale.

We have therefore extended the study of anticipative capability of
a retina from probing it with period stimulations to stochastic stimulations.
Although the responses of the retina induced by these two types of
stimulations seem to be very different, they are of the same nature.
In the OSR, a clear transient, spontaneous (anticipative) response
can be observed in the phenomenon of OSR after the termination of
the periodic stimulations while there seems to be no clear anticipative
responses can be identified after the termination of the stochastic
stimulations. However, the results shown in Fig.~\ref{fig6:latency}
show that the retina is generating signals ahead of the stimulation
with similar information. That is: the retina is actively producing
spontaneous output corresponding to future events of the stimulation;
similar to the case of OSR.

The picture emerges from the above discussions is that the retina
is spontaneously/actively producing output to predict the future.
The location of $\delta t_{p}$ can then be used as an indicator of
the complexity or predictability of the stimulation. Intuitively,
$\delta t_{p}$ will be more negative when the signals is more complex
or more difficult to predict.

\begin{figure}
\begin{centering}
\includegraphics[width=0.9\columnwidth]{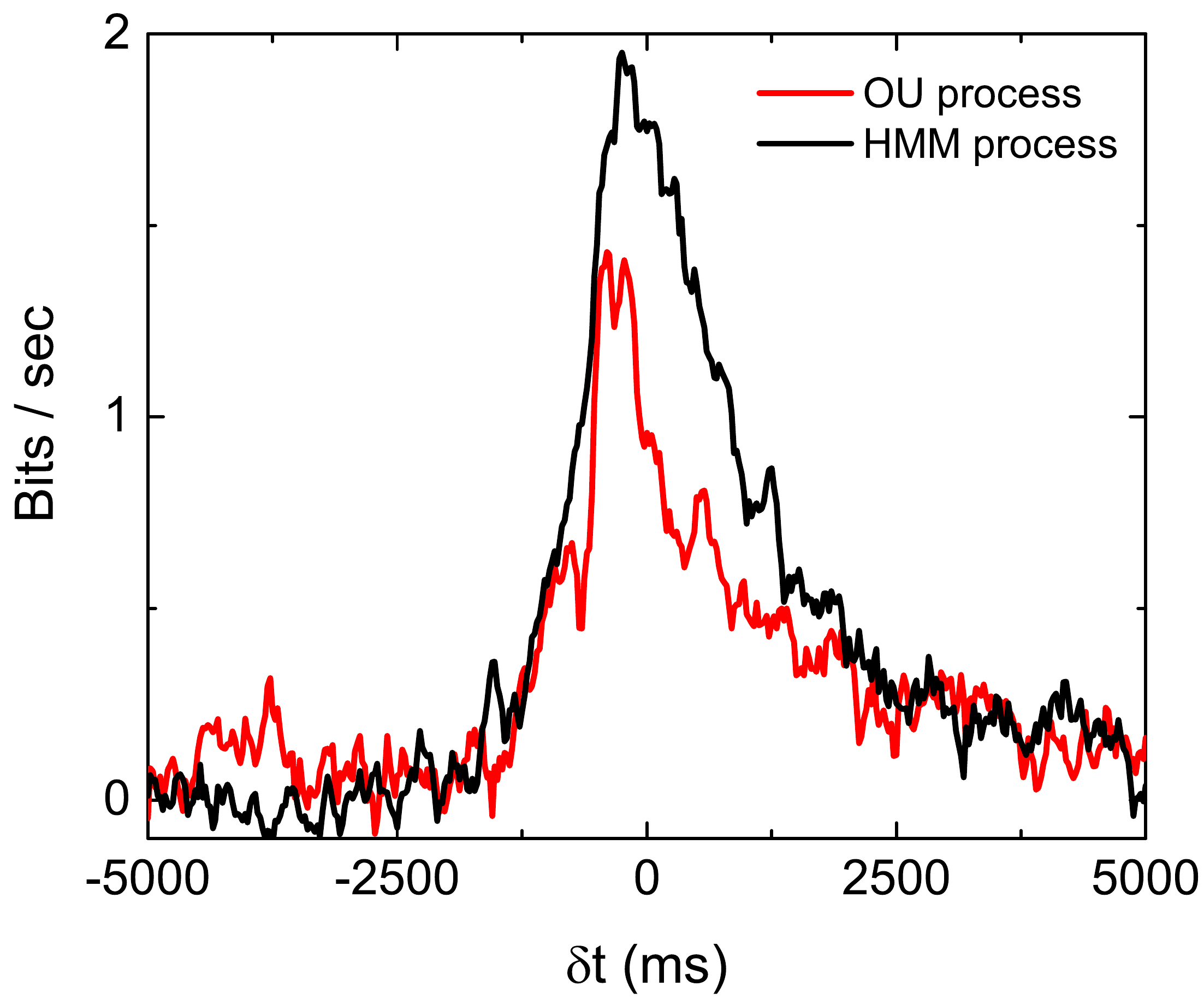} 
\par\end{centering}
\caption{Discriminating OU and the HMM process by a retina. Measured $I_{m}\left(\delta t\right)$
with stimulations generated from an OU process (red) and that similar
to Fig.~\ref{fig1:stochastic pulses}a (black); with $\langle\tau\rangle=200$
ms and $\tau_{cor}=4$ s for both stimulations.}
\label{fig7:discriminating} 
\end{figure}

To test this idea, we have also performed experiments with stimulations
generated by an OU process. We have tuned the OU process in such that
its time scales and fluctuations are similar to those used in our
experiments reported above. Figure~\ref{fig7:discriminating} is
a comparison for $I_{m}$ obtained from the OU process and that from
Fig.~\ref{fig3:comparison}. It can be seen that the peak of $I_{m}$
from the OU process lags behind from that of our experiment; demonstrating
that retina can distinguish these two types of signals and indicate
that the time series from the OU process is more complex that that
used in our experiment. Note that the OU process is a Markovian process
while the time interval signal used in our experiment is not. There
is a hidden variable $v$, which can be deduced from successive values
of pulse interval, forming a hidden Markov Model. Our experimental
results show that the retina somehow manages to make use of this hidden
information to anticipate the next time interval and therefore produce
a peak in $I_{m}$ which can be located at $\delta t_{p}>0$. It is
well known that neural field models for a retina can successfully
produce the anticipative tracking of a moving object spatially. It
is still not known how a retina can do the same in the time domain
\cite{yang_adaptive_2015}. 

\end{document}